\documentclass[a4paper,fleqn,usenatbib]{mnras}

\usepackage[T1]{fontenc}
\usepackage{ae,aecompl}

\usepackage{ulem}
\usepackage{graphicx}

\usepackage{txfonts}

\usepackage{hyperref}
\hypersetup{colorlinks=true, linkcolor=blue, citecolor=blue, urlcolor=blue}

\usepackage{color}
\usepackage{amstext}
\usepackage{multirow}

\newcommand*\lephare{L\textsc{e} P\textsc{hare}}

\newcommand*\Mstar{\mathcal{M}^\star}

\newcommand*\UV{\mathrm{(U-V)}^{0}}
\newcommand*\VJ{\mathrm{(V-J)}^{0}}

\newcommand*\NUVr{\mathrm{(NUV-r)}^{0}}
\newcommand*\rKs{\mathrm{(r-K_s)}^{0}}
\newcommand*\Ks{\mathrm{K_{s}}}

\title[Heavily-obscured X-ray AGNs in slowly quenching $\mathcal{M}^{\star}$ galaxies]{
On the slow quenching of $\Mstar$ galaxies: heavily-obscured AGNs clarify the picture
}

\author[T. Moutard et al.]
{Thibaud Moutard,$^{1}$\thanks{thibaud.moutard@lilo.org}
Nicola Malavasi,$^{2}$
Marcin Sawicki,$^{1}$\thanks{Canada Research Chair}
St\'ephane Arnouts,$^{3}$ and \newauthor 
Shruti Tripathi $^{1}$ \newauthor
\\
$^{1}$Department of Astronomy and Physics and Institute for Computational Astrophysics, Saint Mary's University, 923 Robie Street, Halifax, NS, B3H 3C3, \\
~~ Canada \\
$^{2}$Institut d'Astrophysique Spatiale, CNRS (UMR 8617), Universit\'e Paris-Sud, B\^atiment 121, Orsay, France\\
$^{3}$Aix Marseille Universit\'e, CNRS, LAM - Laboratoire d'Astrophysique de Marseille, 38 rue F. Joliot-Curie, F-13388, Marseille, France
}

\date{Accepted XXX. Received YYY; in original form ZZZ}

\pubyear{2018}

\begin{document}
\label{firstpage}
\pagerange{\pageref{firstpage}--\pageref{lastpage}}
\maketitle



\begin{abstract}
We investigate the connection between X-ray and radio-loud AGNs and the physical properties of their evolved and massive host galaxies, focussing on the mass-related quenching channel followed by $\Mstar (\simeq 10^{10.6} M_\odot)$ galaxies in the rest-frame NUVrK colour diagram at $0.2 < z < 0.5$.
While our results confirm that (1) radio-loud AGNs are predominantly hosted by already-quenched and very massive ($M_*>10^{11}M_\odot$) galaxies, ruling out their feedback as a primary driver of $\Mstar$ galaxy quenching, we found that (2) X-ray AGNs affected by heavy obscuration of their soft X-ray emission are mostly hosted by $\Mstar$ galaxies that are in the process of quenching. 
This is consistent with a quenching scenario that involves mergers of (gas-poor) $\Mstar$ galaxies \textit{after} the onset of the quenching process, i.e., a scenario where $\Mstar$ galaxy mergers are not the cause but rather an aftermath of the quenching mechanism(s).  In that respect, we discuss how our results may support a picture where the slow quenching of $\Mstar$ galaxies happens due to halo-halo mergers along cosmic filaments. 
\end{abstract}






\section{Introduction}
\label{introduction}
The bimodality between blue/star-forming (SF) and red/quiescent (Q) galaxies in rest-frame colour-based diagrams has been extensively documented over the last decades \citep[e.g.,][]{ChesterRoberts1964, Hogg2003, Bell2004, Baldry2006, Arnouts2007, Williams2009, Arnouts2013, Ilbert2013, Bouquin2015, Moutard2016a, Moutard2016b, Pacifici2016a} and the build up of the quiescent population, observed to redshift $z \sim 4$ \citep[e.g.,][]{Tomczak2014, Mortlock2015, Davidzon2017}, is the statistical expression of the seemingly definitive shutdown of the star formation, commonly called \textit{quenching}.
However, the diversity observed among quiescent galaxies (e.g., in terms of mass and morphology) suggests that the mechanisms involved in quenching are multiple, and several studies have stressed the existence of different types of quenching \citep[e.g.,][]{Faber2007, Peng2010, Schawinski2014, Moutard2016b, Pacifici2016b}.

In particular, the fact that star formation is observed to stop earlier in more massive galaxies, on average, underlies the \textit{downsizing} of the quenching \citep[see, e.g.,][whose Figure 15 is edifying]{Bundy2006, Davidzon2013, Moutard2016b}, which argues for the existence of mass-related quenching processes that turn out to have been in operation over a wide range of cosmic times,
given that quiescent galaxies drive the high-mass end of the stellar mass function (SMF) since $z \sim 1.5-2$ \citep[e.g.,][]{Baldry2012, Moustakas2013, ArcilaOsejoSawicki2013, Tomczak2014, Moutard2016b, Davidzon2017}.
Complementarily, the exact Schechter function \citep{Schechter1976} shape of the SMF of star-forming galaxies implies that the probability of quenching increases exponentially with galaxy stellar mass above the characteristic stellar mass $\Mstar$, which brought the idea of \textit{mass quenching} \citep[][]{Ilbert2010, Peng2010}. 
Finally, the fact that $\Mstar$ has actually been found to be very stable for SF galaxies \citep[at least from $z \sim 1$, with $\Mstar = 10^{10.64 \pm 0.01} M_{\odot}$ at $0.2 < z < 1.5$; see][]{Moutard2016b} supports a picture where, on average, galaxies start leaving the star formation main sequence when their stellar mass reaches $\sim\Mstar$, which allows and invites us to consider the quenching of $\Mstar$ galaxies as the natural expression of \textit{mass quenching} at $z \lesssim 1.5$.

Assuming a stellar-to-halo mass relation \citep[e.g.,][]{Coupon2015, Legrand2018}, the characteristic stellar mass of $\Mstar \simeq 10^{10.6}M_{\odot}$ corresponds to a dark-matter (DM) halo critical mass of a few $10^{12} M_{\odot}$. 
While this critical DM halo mass has been shown to be in good agreement with the transition mass above which virial shock heating processes start acting within DM halos \citep[preventing further cold gas accretion; e.g.,][]{Keres2005, DekelBirnboim2006}, other mechanisms able to halt the cold gas supply have been put forth to explain the star formation quenching in massive galaxies. 
In particular, active galactic nucleus (AGN) feedback has been claimed to have an important part in the quenching of massive galaxies, through removal \citep[in the so-called "radiative/cold mode"; e.g.,][]{Hopkins2006, Menci2006} or heating \citep[in the so-called "radio/hot mode"; e.g.,][]{Best2005, Croton2006} of the gas reservoir. 

In that respect, it is interesting that recent studies have shown the quenching of $\sim\Mstar$ galaxies to be a relatively slow process from $z\sim1$, characterized by physical durations of 1-to-a few Gyrs for galaxies to become quiescent after leaving the SF main sequence  \citep[e.g.,][]{Schawinski2014, Ilbert2015, Moutard2016b, Pacifici2016b, Pandya2017}. 
Such observations suggest indeed that $\Mstar$ galaxy quenching is consistent with \textit{starvation} processes, where the cold gas fuelling is impeded and star formation is allowed until the gas reservoir of the galaxy is consumed \citep[e.g.,][]{Larson1980, Peng2015}.
In other words, the quenching of $\Mstar$ galaxies appears to be consistent with heating processes of the gas reservoir, which may result from virial shocks and/or AGN feedback.

While highlighting the connection between the properties of AGNs and their host galaxies is challenging when the rest-frame ultraviolet--to--near-infrared (UV--to--near-IR) part of the spectrum is dominated by the AGN emission (typically the case of Type-1 AGNs), the analysis becomes doable when focussing on optically-obscured AGNs (i.e., with Type-2 AGNs).
The connection between the properties of Type-2 AGNs and the rest-frame optical colours of their host galaxies has been extensively investigated over the past decade \citep[e.g.,][]{Nandra2007, Silverman2008, Georgakakis2008, Bundy2008, Hickox2009, Povic2012, Trayford2016}. 
Of particular interest is the work of \citet{Hickox2009}, who analysed the rest-frame $u$-$r$ colour distribution of X-ray and radio-loud AGN host galaxies as a function of their rest-frame $u$ magnitude to show, in particular, that radio-loud AGNs appear to be predominantly hosted by quiescent galaxies.
Nevertheless, galaxy rest-frame optical colours have been shown to be degenerated, due to the well-known dust reddening, which leads to confusion between quiescent and dusty star-forming galaxies, especially for massive galaxies \citep[][]{Williams2009, Arnouts2013}.

A clearer view of question was given by making use of dust-corrected optical colours, by considering simultaneously the optical and near-IR emission of Type-2 AGN host galaxies, e.g., within the rest-frame U-V vs. V-J colour-colour (hereafter, UVJ) diagram introduced by \citet{Williams2009}, in which (dusty) SF and Q populations are clearly separated, or by using other estimates of the dust attenuation \citep[e.g.][]{Bongiorno2012}. 
For instance, by taking advantage of the UVJ diagram, \citet{Cardamone2010} and \citet{Mullaney2012} could properly assess the incidence of Type-2 X-ray AGNs within Q and SF galaxies. 
None of those studies could, however, highlight any particular trend regarding the interplay between nuclear and star formation activities, as both Q and SF galaxies were found to host Type-2 X-ray AGNs. Moreover, even dust-corrected, optical colours trace fairly old star formation and are therefore not well suited to investigating the variation of the star formation rate in galaxies \citep[e.g.,][]{Martin2007}.

Considering the near-UV emission of galaxies, \citet{Schawinski2014} investigated the distribution of the rest-frame (dust-corrected-) NUV--u--r colours of Type-2 AGN host galaxies, which allowed them to properly show that Type-2 AGNs are hosted by all kinds of galaxies, star-forming,  quiescent and transitioning galaxies. 
However, while such dust-corrected NUV--optical colour-colour diagrams allow, in principle, a proper definition of the so-called green-valley that it is not the result of the confusion between star-forming galaxies and quiescent galaxies but really the region of the colour space frequented by galaxies with intermediate specific star formation rate (sSFR), the diagnostics requires that dust is correctly corrected for, which is notoriously not straightforward without near-IR information, at least, as different extinction laws may apply to identical $E(B-V)$ values \citep[e.g.,][]{Prevot1984, Cardelli1989, Calzetti2000, Arnouts2013}.

In that respect, the rest-frame NUV--r vs. r--K colour-colour (hereafter, NUVrK) diagram which was introduced by \citet{Arnouts2013} is a powerful alternative to the UVJ diagram, as it simultaneously traces recent star formation (thanks to rest-frame near-UV) and dust attenuation (thanks to rest-frame r--K).
This allows the clear identification and a proper definition of the so-called green valley between SF and Q galaxies: in the NUVrK diagram, the green-valley width is $\Delta\NUVr \sim 1$ mag while being unambiguously frequented by galaxies with intermediate sSFR \citep[][]{Arnouts2013,Moutard2016a,Siudek2018}.
The NUVrK diagram is therefore of utmost interest to study galaxies that are in the process of quenching \citep{Moutard2016b}.

In the present paper, we intended to verify whether the quenching of $\Mstar$ galaxies can be related to the presence of AGNs and, notably, to radio-loud AGN feedback.
Taking advantage of the rest-frame NUVrK diagram to identify $\Mstar$ galaxies along the entire quenching channel they follow through the green valley, from the star formation main sequence to the quiescent sequence, we investigate the physical properties of (optically-obscured) AGN host galaxies (the physical parameter estimation of which is not expected to be affected by the AGN emission). 

The paper is organised as follows. In Section~\ref{sect_NUVrK_MsCh}, we present the NUVrK diagram and the associated identification of the quenching channel followed by $\Mstar$ galaxies. In Section~\ref{sect_data}, we present our parent and AGN-host galaxy samples.
We then present our results in Section~\ref{sect_results}, describing the distribution of radio-loud and X-ray AGN host galaxies along the $\Mstar$ quenching channel. We finally discuss the implications of our results for the slow quenching of $\Mstar$ galaxies in Section~\ref{sect_discussion}.

Throughout the paper, we use the standard cosmology ($\Omega_m~=~0.3$, $\Omega_\Lambda~=~0.7$ with $H_{\rm0}~=~70$~km~s$^{-1}$~Mpc$^{-1}$). Magnitudes are given in the $AB$ system \citep{Oke1974} and galaxy stellar masses are given in units of solar masses ($M_{\odot}$) for a \citet{Chabrier2003} initial mass function.

\section{NUVrK diagram and $\Mstar$ quenching channel}
\label{sect_NUVrK_MsCh}


The rest-frame NUV--r vs. r--K (NUVrK) diagram introduced by \cite{Arnouts2013} has been shown to be a powerful alternative to the rest-frame U--V vs. V--J diagram \citep[or UVJ diagram;][]{Williams2009}, as it is sensitive to very different star lifetimes on each of its axes:  $< 0.1$ Gyr along the rest-frame NUV--r colour, rest-frame NUV tracing almost instantaneous star formation rate (SFR) through the emission of massive young stars \citep[typically, O/B stars;][]{Salim2005, Martin2007}; and $>1$ Gyrs along the rest-frame r--K colour, which results from the combination of stellar ageing (the accumulation of low-mass stars) and dust reddening \citep[the UV absorption and IR re-emission of the dust;][]{Arnouts2007, Williams2009}. In other words, while the $\NUVr$ colour varies with specific SFR (sSFR), one can expect the $\rKs$ colour to redden with cosmic time.

Thereby, the NUVrK diagram allows to properly resolve the region of the colour space between quiescent and star-forming galaxies \citep[see][]{Moutard2016a, Moutard2016b, Moutard2018, Siudek2018}, which corresponds to what is often referred to as the \textit{green valley} (GV). 
While the \textit{green valley} has been defined in different ways in the literature (rest-frame optical colour vs. stellar mass, rest-frame optical--to--near-IR colours, ...), the use of the NUVrK diagram really allows us to follow GV galaxies as their sSFR evolves between the SF and Q sequences. 
In other words, in the NUVrK diagram, the green valley is unambiguously frequented by galaxies with intermediate sSFR.
Following \citet{Moutard2016b}, the green valley can be defined as the region of the $\mathrm{NUVrK}$ colour space that satisfies  
\begin{eqnarray}
\bigg[ ~~~~ \left[ ~\NUVr < 3.772 - 0.029 \times t_{\textsc{l}} ~\right] ~ \cup \bigg. ~~~~ \nonumber  \\
\bigg. \left[ ~\NUVr  < 2.25 \times \rKs  + 2.768 - 0.029 \times t_\textsc{l} ~\right] ~~~~ \bigg] ~~ \nonumber\\
\cap ~~~~~~~~
\bigg[  ~~~~ \left[ ~\NUVr > 2.922 - 0.029 \times t_{\textsc{l}} ~\right] ~ \cap  \bigg. ~~~~ \nonumber\\
\bigg. \left[ ~\NUVr  > 2.25 \times \rKs  + 1.918 - 0.029 \times t_\textsc{l} ~\right] ~~~~ \bigg] \ ,
\label{eq_GVsel}
\end{eqnarray}
where $t_\textsc{l}$ is the lookback time in Gyrs.


In addition, the number of degeneracies affecting the NUVrK diagram is quite limited \citep[][]{Moutard2016b, Siudek2018}, which makes this diagram particularly well suited to distinguish star-formation histories (SFHs) characterized by different quenching time-scales \citep[][]{Arnouts2013, Moutard2016b}. 
Provided that the galaxy sample one considers is sufficiently large to probe rare populations, one can isolate the NUVrK pathway that is followed by evolved and massive quenching galaxies on their way to join the quiescent population \citep[80\% of green-valley galaxies are concentrated within the rest-frame colour slice $0.76<\rKs<1.23$;][]{Moutard2016b}.

Typically followed by fairly massive galaxies after reaching a characteristic stellar masses around $\Mstar \simeq 10^{10.64} M_\odot$ (60\% of these green-valley galaxies have stellar masses of $10^{10.5} < M_*/M_\odot < 10^{11}$), this \textit{$\Mstar$ quenching channel} has turned out to be characterised by fairly long quenching timescales, namely, with physical durations of $\sim 1-3.5$ Gyrs to cross the green valley\footnote{As estimated from models of average SFHs with SFR$(t<t_Q) =$ constant and SFR$(t\geq t_Q) \varpropto e^{-(t-t_Q)/\tau_Q}$ where $t$, $t_Q$ and $\tau_Q$ are the cosmic time, quenching time and timescale, respectively \citep[for more detail, please refer to][]{Moutard2016b, Moutard2018}.} as opposed to the quenching of young, $\rKs < 0.76$, low-mass ($M_* \lesssim 10^{9.5}M_\odot$) galaxies that is expected to be 2--9 times faster \citep{Moutard2016b} and to be driven by galaxy local environment \citep{Moutard2018}.

\begin{figure}
\includegraphics[width=\hsize, trim = 0.55cm 0.85cm 0.25cm 0.5cm, clip]{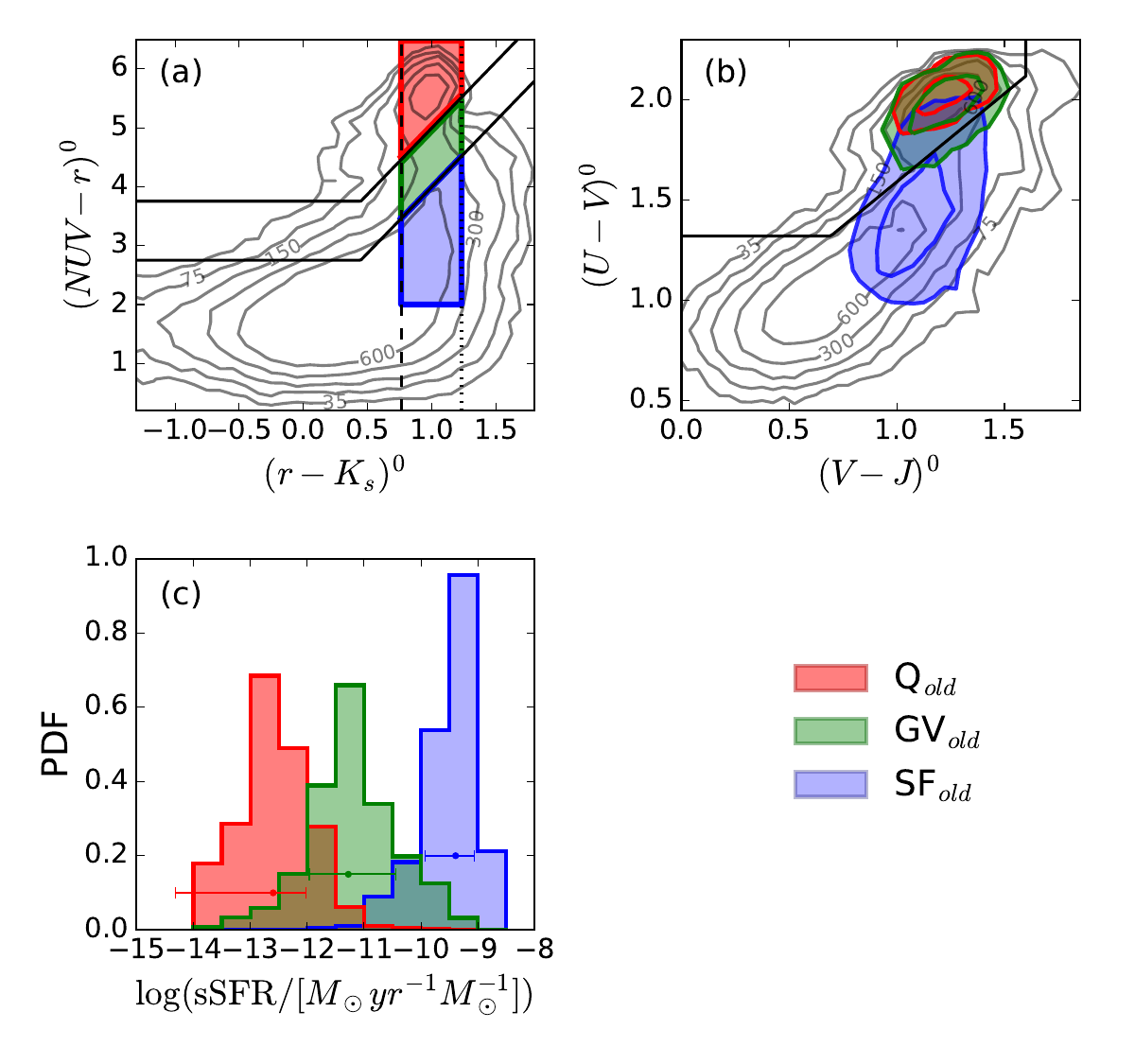}
\caption{
Comparison of the galaxy distribution within the rest-frame $\mathrm{NUVrK}$ and UVJ colour-colour diagrams at $0.2 < z < 0.5$, and identification of green-valley galaxies.
\textbf{a)} NUVrK galaxy distribution and associated identification of the green valley between the SF and Q sequences (as delimited with black solid lines; see Equation~\ref{eq_GVsel}). 
Vertical black dashed and dotted lines delineate the slice of the colour space where 80\% of quenching galaxies concentrate, $0.76 < \rKs < 1.23$ (see Sect.~\ref{sect_NUVrK_MsCh}).
Blue, red and green shaded areas show the NUVrK regions occupied by old, $\rKs > 0.76$ SF, Q and GV galaxies, respectively.
\textbf{b)} UVJ galaxy distribution and associated limit between Q and SF galaxies \citep[black solid line; as identified by][]{Williams2009}.
Blue, red and green filled contours show the distribution of NUVrK-selected old SF, Q and GV galaxies (inner and outer contours including 50 and 90\% of the distribution, respectively). 
\textbf{c)} Corresponding sSFR probability distribution functions (PDFs). Points and errorbars show the median and the 16\% and 84\% percentiles of the PDFs.
\label{fig_NUVrKvsUVJ}    }
\end{figure}

To illustrate the interest of the $\mathrm{NUVrK}$ diagram to analyse the quenching of $\Mstar$ galaxies, with respect to the $\mathrm{UVJ}$ diagram, we compare in Fig.~\ref{fig_NUVrKvsUVJ} the distribution of galaxies drawn from our parent galaxy sample (see below, Sect.~\ref{sect_phys_param}).
Fig.~\ref{fig_NUVrKvsUVJ}a shows the NUVrK galaxy distribution and the associated identification of the green valley, as defined in Equation~\ref{eq_GVsel}.
In particular, that allows us to distinguish old, $\rKs > 0.76$ intermediate-sSFR galaxies of the green valley (with sSFR~$\sim10^{-10.5-12} M_\odot \mathrm{yr}^{-1} M_\odot^{-1}$; green) from really quiescent galaxies (with sSFR~$<10^{-12} M_\odot \mathrm{yr}^{-1} M_\odot^{-1}$). 
The corresponding sSFR probability distribution functions shown in Fig.~\ref{fig_NUVrKvsUVJ}c confirm indeed that NUVrK-selected, old GV galaxies are characterized by a range of sSFR that is clearly different from those of Q and SF galaxies (namely, only only 16\% of old, NUVrK-selected GV and Q galaxies may have similar sSFR, and less than 16\% of old, NUVrK-selected GV and SF galaxies may have similar sSFR).
In contrast, the UVJ distribution of the very same galaxy sample shown in Fig.~\ref{fig_NUVrKvsUVJ}b does not allow such distinction, as NUVrK-selected GV and Q galaxies have similar UVJ colours \citep[which has already been shown by][Figures 2a and 2c]{Siudek2018}. This is particularly true for old, $\rKs > 0.76$ galaxies,\footnote{Note that $\rKs = 0.76$ is about $\VJ \sim 1$ around the green valley.} of interest in the following analysis, with more than 80\% of NUVrK-selected, old GV galaxies having UVJ colours that are undistinguishable from the UVJ colours of old Q galaxies.
In other words, this confirms that $\UV$ is a poor tracer of the sSFR for low- and intermediate-sSFR galaxies, compared to $\NUVr$.

\section{Parent and AGN-host galaxy samples}
\label{sect_data}

Aiming to analyse the connection between AGNs and the physical properties of their host galaxies, we have focused on AGN host galaxies whose rest-frame UV--to--near-IR emission is dominated by stars (to allow the estimation of physical properties from purely stellar population synthesis models) while the  AGN emission had been observed in extreme wavelength regimes, i.e., galaxies that are hosting an optically-obscured (or Type 2) AGN.

For this purpose, we made use of the comprehensive photometric coverage of the XMM Large-Scale Structure \citep[XMM-LSS;][]{Pierre2004} field, which benefits from homogeneous overlap between UV--to--near-IR broad-band photometry, X-ray and radio observations over $\sim7$ deg$^2$.
Specifically, the parent sample of galaxies and associated physical properties (absolute magnitudes, stellar mass, star formation rate...) has been based on UV--to--near-IR observations while hosted optically-weak AGNs were identified by cross-matching with overlapping X-ray and radio-loud AGN public samples, as detailed in the following section.

\subsection{UV--to--near-IR observations and physical parameters of the parent galaxy sample}
\label{sect_phys_param}

The data we used combine photometry from the VIPERS Multi-Lambda Survey \citep[VIPERS-MLS: FUV, NUV, u-, g-, r-, i-, z- and $\Ks$-bands; for more details, see][]{Moutard2016a} and VISTA Deep Extragalactic Observations \citep[VIDEO: Y-,J-,H-,Ks-bands;][]{Jarvis2013} over $\sim7$ deg$^2$ in the XMM-LSS field, which overlap with the Canada-France-Hawaii Telescope Legacy Survey (CFHTLS) field W1. Photometric redshifts (photo-zs) and physical parameters were derived using the template-fitting code \lephare \ \citep{Arnouts2002, Ilbert2006}, as described in \citet[][]{Moutard2016a, Moutard2016b}. 

Of interest for the present study is the photo-z accuracy, $\sigma_{z} \leq 0.03 (1+z)$ down to $K_s \sim 22$ (i.e., for any galaxy with stellar mass $M_* \geq 10^{9}M_\odot$ and thus for the $\sim \Mstar$ galaxies we considered in this paper).
The careful identification of galaxies from stars and Type-1 AGNs was based on the combination of several diagnostics, notably based on the surface brightness of the sources and comparison of their observed NUV--to--Ks photometry with empirical (i.e. observed) template libraries of stars and AGNs \citep[for more details, see][]{Moutard2016a,Moutard2016b}.

Finally, great care has been taken in the derivation and the use of absolute magnitudes, to minimize the impact of the k-correction: while rest-frame magnitudes were first computed using the nearest observed-frame band, we have limited our analysis to $z < 0.5$ to limit the k-correction on rest-frame $K_s$ magnitude computed from $K_s$ observations (our reddest band).\footnote{All rest-frame magnitudes were derived with two different template libraries, which allowed us to verify, in particular, that our rest-frame $K_s$ magnitude estimates are consistent to $0.2 \leq z \leq 0.5$.}

\subsection{Radio-loud and X-ray AGN properties}
\label{sect_AGN_samples}

Aiming to identify the X-ray and radio-loud AGN host galaxies and explore associated properties, we made use of public catalogues of X-ray and radio-loud AGNs in the XMM-LSS/CFHTLS-W1 field.

\subsubsection{Radio-loud and X-ray AGN catalogues}

While associating X-ray and radio sources with optical counterparts can be challenging, we made use of well-documented photometric catalogues where the determination of X-ray and radio sources optical counterparts has been done.

Regarding radio-loud AGNs, we made use of the catalogue produced by \citet{Tasse2008a}, where radio-loud AGNs were identified based on multi-frequency radio observations with flux density limits of $\sim4$ and $\sim1.5$ mJy/beam at 325 and 610 MHz, respectively \citep[i.e., complete for radio powers $P_\mathrm{1.4GHz} \gtrsim 6 \times 10^{23}$ W/Hz at $z<0.5$; see also ][]{Tasse2006, Tasse2007, Tasse2008b},
and carefully associated to their optical counterparts from the CFHTLS. More specifically, we selected radio-loud AGNs with high optical counterpart association probability ($P^{\prime}  \geq 80$ per cent) and no significant contribution from Type-1 AGN or starburst activity \citep[for more details, see][]{Tasse2008a}.

As for X-ray AGNs, we made use of the catalogue of X-ray AGNs and associated X-ray properties of \citet{Melnyk2013}, with flux limits of $3 \times 10^{-15}\mathrm{erg\ s^{-1} cm^{-2}}$ and $1 \times 10^{-14}\mathrm{erg\ s^{-1} cm^{-2}}$ in the soft [0.5--2 keV] and hard [2--10 keV] bands, respectively (i.e., complete for X-ray luminosities $L_X \gtrsim 0.7\times10^{42}\mathrm{erg\ s^{-1}}$ or $L_X \gtrsim 2\times10^{42}\mathrm{erg\ s^{-1}}$ at $z<0.5$ in the soft and hard bands, respectively). 
The AGN identification was based on the observation of X-ray point-like source from the XMM-LSS survey, combined with their optical counterparts from the CFHTLS \citep[we selected sources classified as "AGN" and/or "QSO"; for more details, see][]{Chiappetti2013, Melnyk2013}.

\subsubsection{Matching of AGN and host galaxy physical properties}

Radio-loud and X-ray AGN host galaxy properties were recovered from position matching with their NUV--to--K counterparts from the parent sample (Sect.~\ref{sect_phys_param}), by using a $0.5^{\prime\prime}$ tolerance radius.\footnote{Some physical parameters are available in XMM-LSS catalogues, but those were derived from early releases of the CFHTLS (namely, T0002 and T0004 for radio-loud and X-ray AGN host galaxies, while the VIPERS-MLS is based on T0007) without homogeneous NUV--to--K photometry.} 
Doing so, we could take advantage of the careful identification galaxies, stars and unobscured AGNs performed in the VIPERS-MLS, in order to ensure that we only considered (optically-obscured AGN host) galaxies, the NUV--to--K emission of which is expected to be dominated by the content in stars and dust.


Aiming to quantify the fraction of galaxies hosting an AGN (see Sect.~\ref{sect_results}), we defined the control galaxy samples associated with radio and X-ray observations: namely, the radio and X-ray control samples were selected as the sub-samples (of the parent sample) lying within the footprint of radio and X-ray observations, respectively.
Thereby, at $0.2 < z < 0.5$, we obtained 28 radio-loud AGN host galaxies in a parent population of 24386 galaxies and 117 X-ray AGN host galaxies in a parent population of 24619 galaxies, within effective areas of 6.51 deg$^2$ and 6.45 deg$^2$, respectively.

\subsection{Optically-obscured (Type-2) AGN host properties}
\label{sect_AGN_samples}

\subsubsection{Securing of our selection of Type-2 AGNs}
\label{sect_Typ2_sel}

While in the present analysis, stars and optically-unobscured (Type-1) AGNs were carefully discarded from the parent galaxy sample (Sect.~\ref{sect_phys_param}), a few Type-1 AGNs may have been missed. 

However, due to the emission of the accretion disc, Type-1 AGNs have very blue rest-frame NUV--r colours. While, most of the time, the fitting of such photometry with galaxy templates fails, very blue/star-forming galaxy templates can fit Type-1 AGN photometry, which then results into very specific (and artificial) rest-frame NUV-r vs. r-K colours. 
\citet{Siudek2018} showed indeed that broad-line AGNs tend to concentrate in the region of the NUVrK diagram with $\NUVr \lesssim 1$ and $\rKs \lesssim 1$, that is in a very blue part of the NUVrK diagram.  

Thus, by considering the region of the $\Mstar$ quenching channel, namely, for galaxies with $0.76 < \rKs < 1.23$, and galaxies with $\NUVr \gtrsim 2$ at $0.2 < z < 0.5$, we exclude the region of the colour space that is prone to be contaminated by Type-1 AGNs. 
We finally verified that all the optically-obscured AGN host galaxies we considered in the following analysis were always best fitted with a galaxy template than with a Type-1 AGN template, i.e., $\chi_{GAL}^2~/~\chi_{AGN}^2<1$.

The rest-frame r-band emission (and thus, the stellar mass estimation) of our samples of X-ray and radio-loud AGN host galaxies is therefore not expected to be affected by the AGN emission, .

\subsubsection{On the near-UV and near-IR contribution of Type-2 AGNs to their host emission}
\label{sect_AGN_NUVK_contrib}

By discarding Type-1 AGNs and considering only Type-2 AGNs (see Sect.~\ref{sect_Typ2_sel}), we only considered optically-obscured AGNs whose rest-frame contribution in the r-band is negligible.
On the other hand, our analysis (see Sect.~\ref{sect_results}) relies on the assumption that our measurements of galaxy rest-frame NUV-r and r-K colours are not significantly affected by the emission of hosted AGNs. The potential contribution of Type-2 AGNs to the rest-frame emission of their host galaxy in the near-UV and near-IR  has therefore to be assessed.

\begin{figure}
\includegraphics[width=\hsize, trim = 0.cm 0.2cm 0.cm 0.5cm, clip]{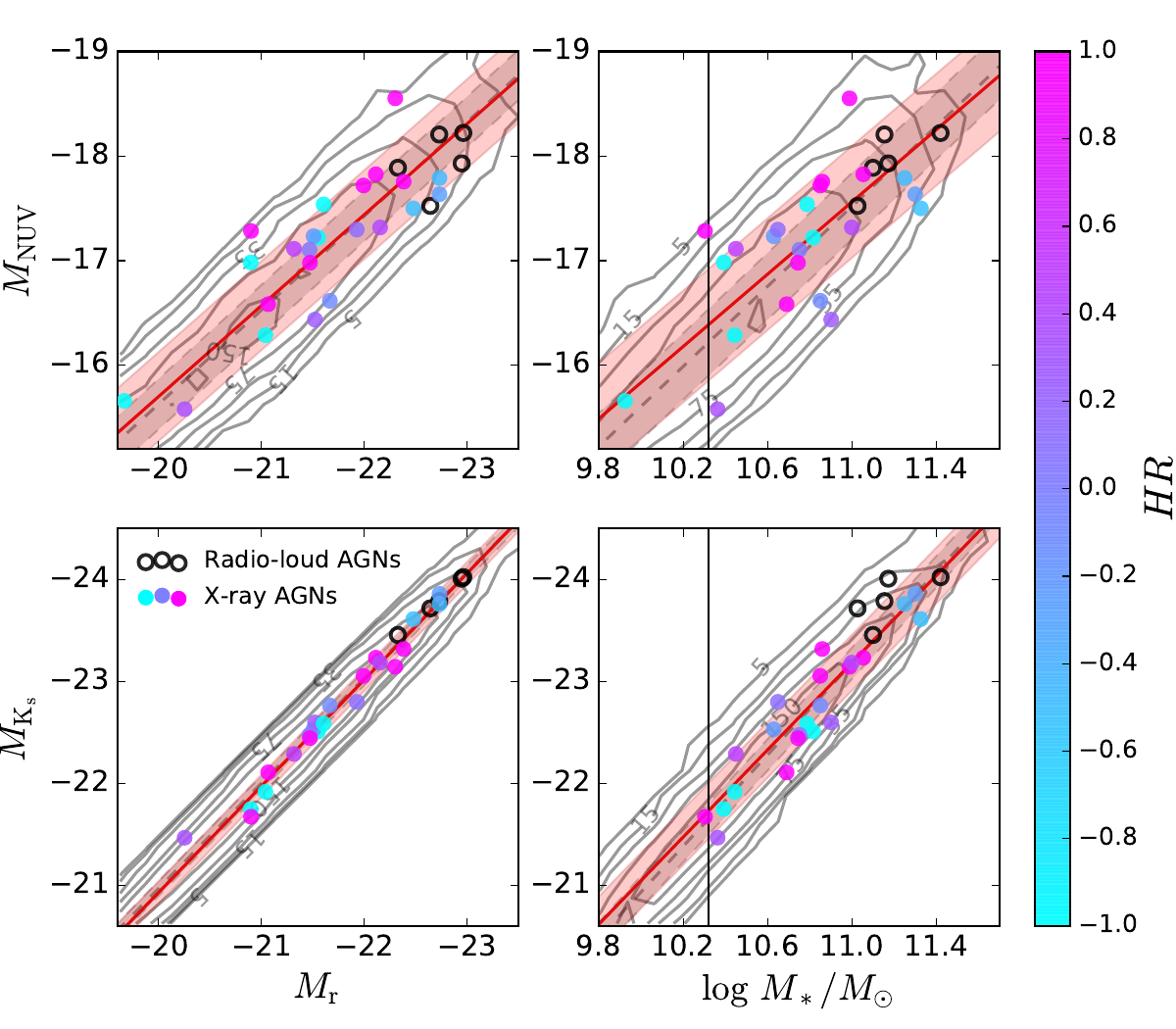}
\caption{
NUV and K-band absolute magnitudes of optically-obscured X-ray (colour-coded filled circles) and radio-loud (open black circles) AGN host galaxies in the green-valley, as a function of r-band absolute magnitude and stellar mass, with respect to the distribution of the parent sample of galaxies (grey contours).
The linear regressions (and corresponding $\pm1\sigma$ dispersions) of the Type-2 AGN host and parent galaxy samples are shown with the oblique red solid and grey dashed lines (and associated envelopes), respectively.
The stellar mass completeness limit we adopted when comparing the different samples (see Sect.~\ref{sect_AGN_NUVK_contrib}) is shown with black solid line.
\label{fig_r_Mass_vs_NUV_K}    }
\end{figure}

As described in Section \ref{sect_NUVrK_MsCh} and observed in Fig.~\ref{fig_NUVrKvsUVJ}, intermediate-sSFR galaxies are remarkably concentrated in the green valley of the NUVrK diagram, 80\% of them having rest-frame colours $0.76 < \rKs < 1.23$. 
Precisely associated with the quenching channel of $\Mstar$ galaxies which we investigate in the present paper (cf. Sect.~\ref{sect_results}), this fairly narrow region of the NUVrK colour space is the result of the secular evolution of the stellar- and dust-content within galaxies which, on average, saturates around $\rKs \sim 1$ while extremely red $\rKs \gtrsim 1.2$ colours are only possible for edge-on galaxies \citep[cf. Sect. \ref{sect_NUVrK_MsCh}; see also][]{Arnouts2013, Moutard2016a}.
Any significant rest-frame K-band emission of Type-2 AGNs would therefore tend to redden the $\rKs$ colour distribution of their host galaxies, while any significant emission of Type-2 AGNs in the NUV would to make the $\NUVr$ bluer.

Fig.~\ref{fig_r_Mass_vs_NUV_K} shows the rest-frame NUV (top) and K$_s$ (bottom) magnitude distribution of green-valley optically-obscured (Type-2) AGN host galaxies as a function of the rest-frame $r$-band magnitude (left) and stellar mass (right), along with the associated distribution of the parent sample of galaxies (grey contours).
Overall, one can see that the stellar mass and rest-frame NUV--r--K$_s$ distribution of Type-2 AGN hosts follows the distribution of the parent sample of galaxies.
More specifically, at given $r$-band absolute magnitude or stellar mass, the divergence that is observed between the average NUV and K$_s$ absolute magnitudes of galaxies of the parent sample  (grey dash-dotted lines) and Type-2 AGN host galaxies (red dashed lines) is always smaller than the dispersion associated with the average NUV and K$_s$ absolute magnitudes (red and grey envelopes).
In other words, there is no particular enhancement of the NUV and K$_s$ emission of Type-2 AGN host galaxies with respect to non-AGN galaxies.

This tends to confirms that, in our sample of optically-obscured AGN host galaxies, the AGN contribution to rest-frame NUV and K$_s$ emission is negligible.

\subsubsection{Stellar mass completeness of our parent and  X-ray and radio-loud AGN host galaxy samples}
\label{sect_sample_comp}

\begin{figure}
\includegraphics[width=0.96\hsize, trim = -0.5cm 0cm 0cm 0cm, clip]{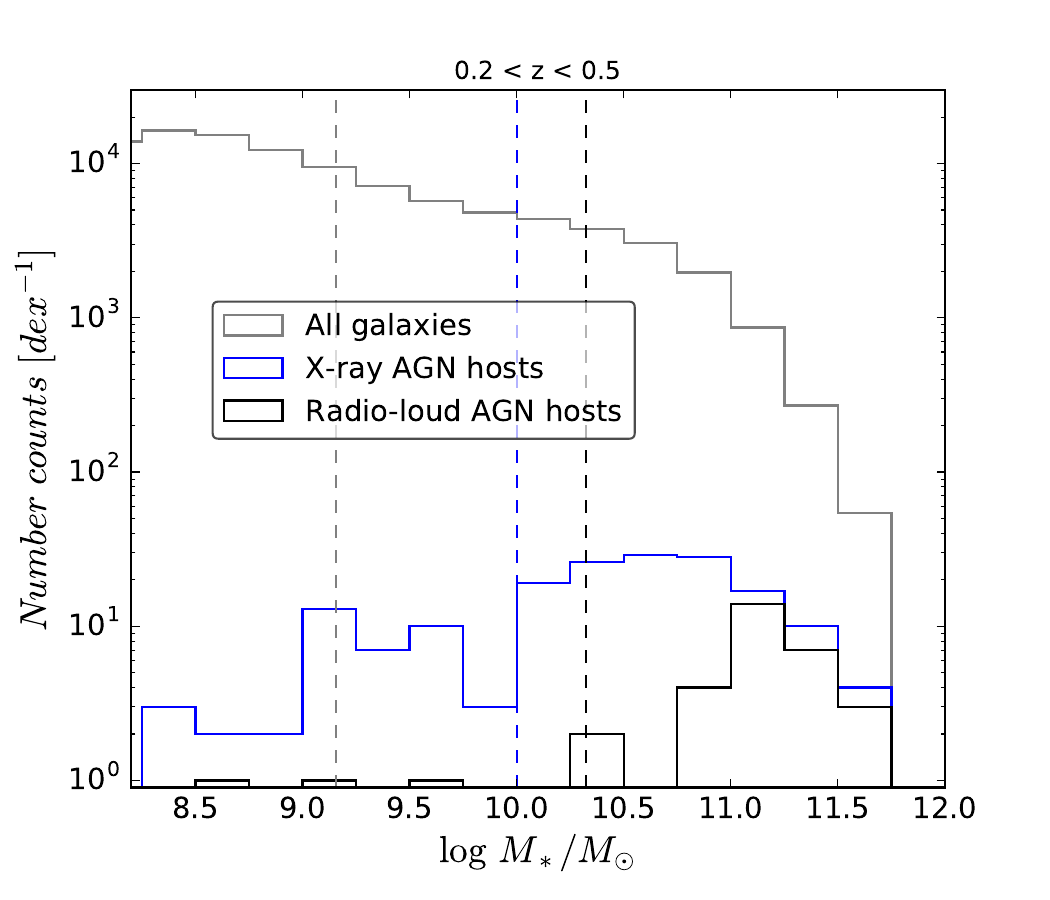}
\caption{
Stellar mass distribution X-ray (blue solid line) and radio-loud (black solid line) AGN host galaxies, compared with that of the parent galaxy sample (grey solid line) at $0.2 < z < 0.5$.
Vertical dashed lines show the associated stellar mass completeness limits at $z \simeq 0.5$, with $\log(M^{gal}_\mathrm{lim}/M_\odot) = 9.16$, $\log(M^{Xray}_\mathrm{lim}/M_\odot) = 10.00$ and $\log(M^{Radio}_\mathrm{lim}/M_\odot) = 10.32$, the limits of parent, X-ray AGN host and radio-loud AGN host samples, respectively (see Sect.~\ref{sect_sample_comp}).
\label{fig_AGNhost_MassDistrib}    
}
\end{figure}

Fig.~\ref{fig_AGNhost_MassDistrib} shows the stellar mass distribution of X-ray and radio-loud AGN host galaxies (Sect.~\ref{sect_AGN_samples}), compared to the distribution of the parent sample of galaxies (Sect.~\ref{sect_phys_param}). 
Following \citet{Pozzetti2010}, we derived the stellar mass completeness limit from the  distribution of the lowest stellar mass at which a galaxy could be detected given its redshift, $M_{min}$. 
If the sample is limited by the observed flux $f$ (or magnitude, $m$), down to the limiting flux $f_{\mathrm{lim}}$ (or magnitude $m_{\mathrm{lim}}$), $M_{min}$ is given by
\begin{equation}
\log(M_{min}) = \log(M_*) - \log \left(  \frac{f}{f_{\mathrm{lim}}} \right)  ~,
\label{eq_flux_mass_lim}
\end{equation}
or, in terms of magnitude,
\begin{equation}
\log(M_{min}) = \log(M_*) + 0.4 \ (m -m_\mathrm{lim})  ~.
\label{eq_mag_mass_lim}
\end{equation}
We conservatively considered the 20\% highest redshift galaxies (i.e., those that are closest to the upper limit of the redshift bin, $z = 0.5$).
The stellar mass completeness limit, $M_\mathrm{lim}$, is then given by the stellar mass for which 90\% of the population have $M_* > M_{min}$.

For the parent galaxy sample, the limiting magnitude $K_s = 22$ translates into a stellar mass completeness limit $M^{gal}_\mathrm{lim} = 10^{9.16} M_\odot$ at $z \lesssim 0.5$.
Regarding our samples of radio-loud and X-ray AGNs, the selection combines limiting fluxes in two different wavelength ranges, by including any AGN that was detected in one and/or the other band (soft-or-hard X-ray bands and 325-or-610 MHz radio bands, respectively).
In other words, an AGN was detected as long as it was bright enough in any of the two bands.
For X-ray AGN host galaxies, the stellar mass completeness limit can therefore be written as
\begin{equation}
M^{X ray}_\mathrm{lim} = \min_b (M^b_\mathrm{lim})~, ~~ \mathrm{for} ~b = \mathrm{soft,hard}   ~
 \label{eq_Mabs_lim_eff}
\end{equation}
which, given X-ray limiting fluxes of $3 \times 10^{-15} \mathrm{erg\ s^{-1} cm^{-2}}$ in the soft band and $1 \times 10^{-14} \mathrm{erg\ s^{-1} cm^{-2}}$ in the hard band, gives us $\log( M^{X ray}_\mathrm{lim}/M_\odot) =10.00$ at $z \lesssim 0.5$.
Similarly, the  stellar mass completeness limit of radio-loud AGN host galaxies can be written as
 \begin{equation}
 \label{eq_Mabs_lim_eff}
M^{Radio}_\mathrm{lim} = \min_b (M^b_\mathrm{lim})~, ~~ \mathrm{for} ~b = \mathrm{325MHz,610MHz} ~
\end{equation}
which, with limiting flux densities of $4$ mJy/beam at 325 MHz and $1.5$ mJy/beam at 610 MHz, translates into $\log(M^{Radio}_\mathrm{lim}/M_\odot)=10.32$ at $z \lesssim 0.5$.

When comparing the different samples, we only considered galaxies with stellar masses greater than the most restrictive stellar mass completeness limit, which tuns out to be that of the radio-loud AGN host sample, $M_\mathrm{lim} = M^{Radio}_\mathrm{lim} = 10^{10.32} M_\odot$. 
Doing so, we ensured that the comparison between non-AGN galaxies and X-ray and radio-loud AGN host galaxies is sensible.
At the same time, by considering galaxies with stellar masses $\log(M_*) > 10.32$ we could follow galaxies from $\NUVr \sim 2$ up to $\NUVr \sim 6$, i.e., along the whole quenching channel followed by $\Mstar$ galaxies (cf. Section \ref{sect_NUVrK_MsCh}).


\begin{figure}
\center
\includegraphics[width=\hsize, trim = 0.1cm 0cm 0.2cm 0.35cm, clip]{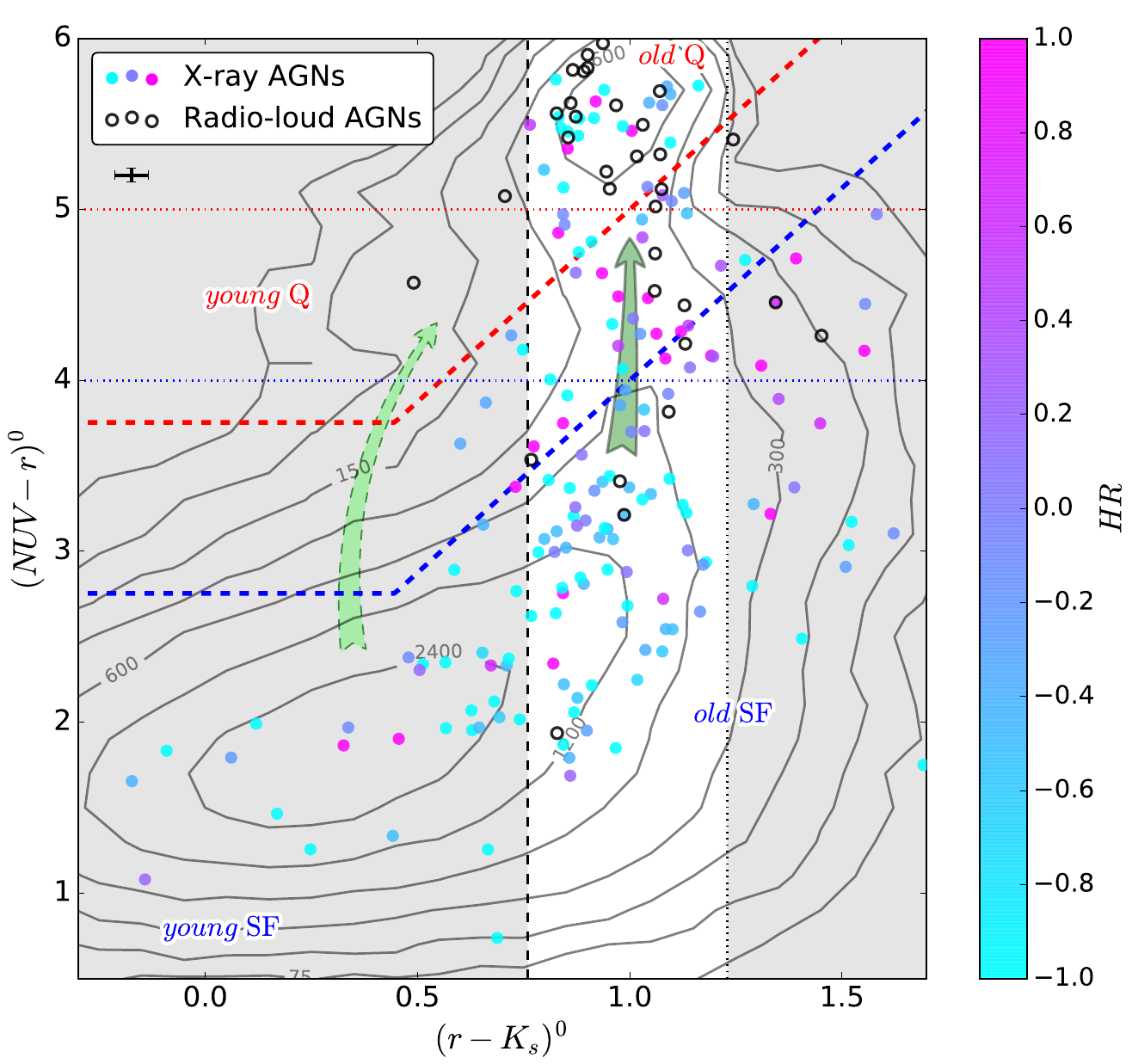}
\caption{NUVrK distribution of AGN host galaxies at $0.2 < z < 0.5$. The distribution of radio-loud (black open circles) and X-ray (color coded filled circles) AGN host galaxies (colour coded according to their hardness ration, $HR$) is compared to the distribution of parent sample (grey contours). The typical uncertainty affecting NUVrK colours is shown in the upper left corner.
Vertical black dashed and dotted lines delineate the slice of the colour space associated with the $\Mstar$ quenching channel and grey shaded regions indicate the regions excluded from our analysis (see Sections \ref{sect_NUVrK_MsCh} and \ref{sect_results}). 
Blue and red dashed lines separate \textit{green valley} galaxies from star-forming (SF) and quiescent (Q) galaxies, respectively (see Equation~\ref{eq_GVsel}).
Horizontal blue and red dotted lines indicate the corresponding mean lower and upper $\NUVr$ limits of the green valley in the $\Mstar$ galaxy quenching channel.
\label{fig_NUVrK}  }
\end{figure}

\begin{figure}
\center
\includegraphics[width=\hsize, trim = 0cm 1.3cm 0cm 1.cm, clip]{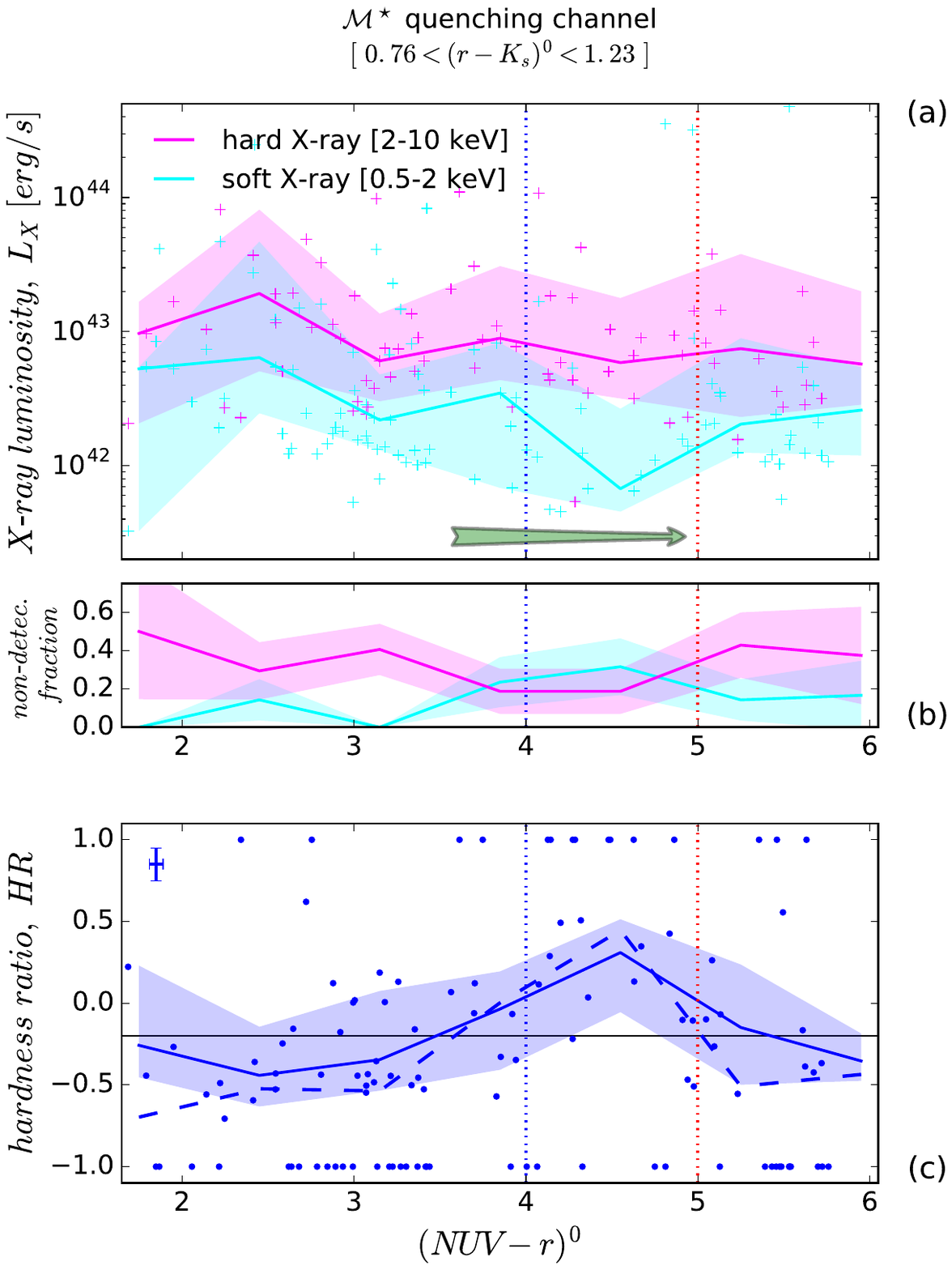}
\caption{X-ray properties of AGN host galaxies with the $\Mstar$ quenching channel.
\textbf{a)} hard [2--10 keV] (magenta) and soft [0.5--2 keV] (cyan) X-ray luminosity ($L_X$), \textbf{b)} corresponding fractions of hard and soft X-ray non-detections, and \textbf{c)} associated hardness ratio are plotted as a function of the $\NUVr$ colour along the $\Mstar$ quenching channel (i.e., for host galaxies with $0.76 < \rKs < 1.23$). 
Median values are shown with solid lines. Shaded envelopes show the corresponding $\pm1 \sigma$ uncertainties (as derived from 10000 bootstrap resamples in panels a, from the Poissonian standard deviation in panel b, and from 1000 mock catalogues perturbed according to the typical uncertainty affecting $HR$ and $\NUVr$ in panel c, shown in the upper left corner). 
Vertical blue and red dotted lines show the mean lower and upper $\NUVr$ limits  associated with the green valley, respectively, within the $\Mstar$ quenching channel (the direction of which is illustrated by the green arrow in panel a).
In panel c, the dashed line shows the median HR derived when including non-detections ($HR=\pm1$); the horizontal black solid line shows the threshold $HR=-0.2$ used to isolate heavily-obscured (soft X-ray) AGNs.
\label{fig_NUVr_Xray}  }
\end{figure}

\begin{figure}
\includegraphics[width=1.05\hsize, trim = 0cm 0.35cm 0cm 0.cm, clip]{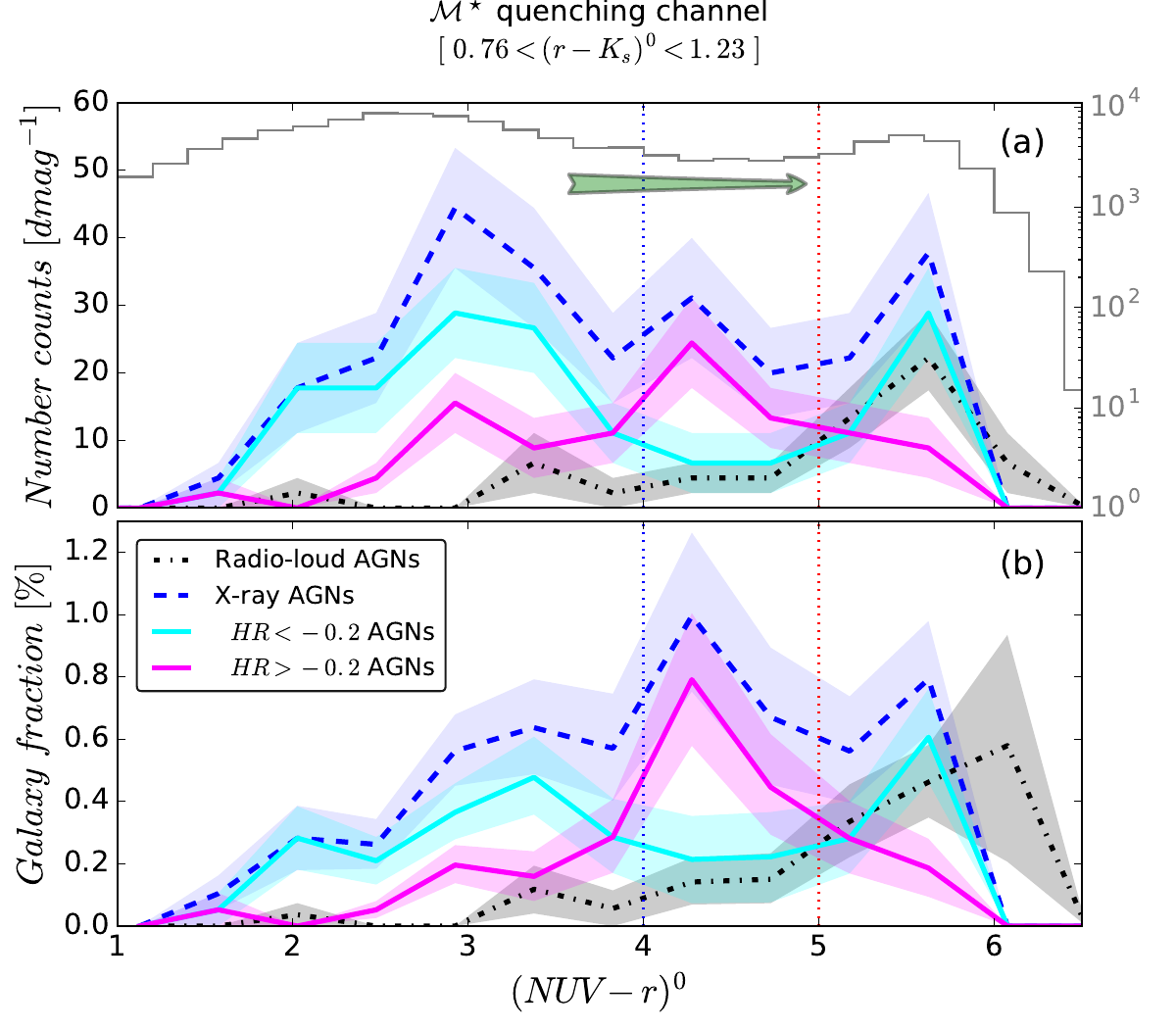}
\caption{$\NUVr$ distribution within the $\Mstar$ quenching channel (the direction of which is illustrated by the green arrow), for radio-loud (black) and X-ray (blue) AGN host galaxies, including $HR<-0.2$ (cyan) and heavily-obscured, $HR>-0.2$ (magenta) X-ray AGNs.
The vertical blue and red dotted lines show the mean lower and upper limits of the green valley (cf. Fig.~\ref{fig_NUVrK}).
\textbf{a)} Number counts of AGN host galaxies at given $\NUVr$ (the distribution of all galaxies from the X-ray control sample is shown in light grey for comparison).
\textbf{b)} Corresponding fraction of galaxies hosting an AGN at given $\NUVr$.
Median values are shown with solid lines while shaded envelopes show corresponding $\pm1 \sigma$ uncertainties (as derived from 10000 bootstrap resamples). Only galaxies with $M_* > M_\mathrm{lim} = M^{Radio}_\mathrm{lim}$ are considered (see Sect.~\ref{sect_sample_comp}).
\label{fig_AGN_NUVr}    }
\end{figure}


\section{Results}
\label{sect_results}

In Fig.~\ref{fig_NUVrK}, we show the distribution of radio-loud and X-ray AGN host galaxies over the distribution of galaxies in the NUVrK diagram (grey contours). 
One can clearly identify the quenching channel (marked with dark green arrow) followed by $\sim \Mstar$ galaxies in the \textit{green valley} when they leave the SF main sequence around $\NUVr \sim 4$ (i.e., sSFR $\sim 10^{-10.4} M_\odot \mathrm{yr}^{-1}/ M_\odot$) to eventually join the quiescent population around $\NUVr \sim 5$ (i.e., sSFR $\sim 10^{-12.4} M_\odot \mathrm{yr}^{-1}/ M_\odot$), all concentrated in a slice of the colour space with $0.76 < \rKs < 1.23$.
Doing so, we ensured that galaxies following the $\Mstar$ quenching channel were isolated from \textit{young} galaxies with $\rKs < 0.76$ \citep[prone to a much faster quenching; light green arrow; see][]{Moutard2016b}, and from very dusty SF galaxies with $\rKs > 1.23$, the $\NUVr$ colour of which is strongly affected by dust absorption \citep[][]{Moutard2016a,Moutard2016b}.
In the following, we focused on this $0.76 < \rKs < 1.23$ slice of the colour space, aiming to study the incidence and the properties of radio-loud and X-ray (optically-obscured) AGN host galaxies along this $\Mstar$ quenching channel.

\subsection{NUVrK colours of Type-2 AGN host galaxies}
\label{sect_NUVrK_AGNs}

The first remark emerging from Fig.~\ref{fig_NUVrK} is that radio-loud and X-ray AGNs appear to be two distinct populations, essentially hosted by different galaxies.
More specifically, while radio-loud AGNs appear to be predominantly hosted by old quiescent galaxies, X-ray AGNs are observed in all kinds of old, $\rKs>0.76$ galaxies over the entire $\NUVr$ range, following the distribution of galaxies. 
However, when considering the hardness of X-ray AGN emission, i.e., the prevalence of hard [2--10keV] over soft [0.5--2keV] X-ray emission, high hardness ratios ($HR$ \footnote{$HR = (H-S)/(H+S)$, where $H$ and $S$ are the numbers of hard [2--10keV] and soft [0.5--2keV] X-ray photons, respectively. In the X-ray catalogue we used, the typical uncertainty on $HR$ is 0.1 \citep{Melnyk2013}. }) appear to be favoured in the green valley.

In Fig.~\ref{fig_NUVr_Xray} we show how the X-ray properties depend on an object's position along the the $\Mstar$ quenching channel.  In these three panels quenching galaxies move from left to the right in $\NUVr$, and the mean lower and upper $\NUVr$ limits of the green valley are marked with blue and red dotted lines at $\NUVr=4$ and 5, respectively. 

Fig.~\ref{fig_NUVr_Xray}a shows the AGN X-ray luminosity as a function of the $\NUVr$ colour of its host galaxy in the hard and soft bands. The hard X-ray emission appears to be stable over the entire $\NUVr$ range (i.e., before, in and after the green valley),  
with a median luminosity of $L^{hard}_X = 8^{+12}_{-4}\times10^{42}$erg/s.  In contrast, the mean soft X-ray emission 
evolves along the quenching channel: it is $L^{soft}_X = 3.5^{+4.5}_{-2.8}\times10^{42}$ erg/s in SF, decreases to $L^{soft}_X = 0.7^{+2.3}_{-0.3}\times10^{42}$ erg/s in the green valley, before recovering in Q galaxies to $L^{soft}_X \sim 3^{+3}_{-2}\times10^{42}$ erg/s -- a level comparable to that of SF galaxies. 
Consistent with this, the median fraction of soft X-ray non-detections increases drastically in the green valley, rising from $\sim$10 per cent in SF galaxies to more than $\sim$40 per cent in the green valley, and then decreasing in Q galaxies down to $\sim$20 per cent, as shown in Fig.~\ref{fig_NUVr_Xray}b.

Indeed, if we conservatively exclude non-detections (i.e., when $HR=\pm1$), this results in a clear increase of the median hardness ratio to $HR = 0.31$ in the green valley, against $HR \simeq -0.4$ in SF and Q galaxies, as observed in Fig.~\ref{fig_NUVr_Xray}c. 
The analysis of the $HR$ distribution for SF, GV and Q galaxies through Anderson-Darling \citep{AndersonDarling1952, AndersonDarling1954} tests confirmed that GV X-ray AGN host galaxies are drawn from different populations than Q and SF X-ray AGN host galaxies (with $>95\%$  and $>99.9\%$ confidence, respectively), while we cannot exclude that Q and SF X-ray AGN host galaxies have similar $HR$ distributions.
Note that if we include $HR=\pm1$ AGNs (i.e., AGNs where not detected in one of the two bands), the trend is even stronger, with $HR = 0.44$ in the middle of the green valley against $HR \simeq -0.5$ outside (dashed blue line in Fig.~\ref{fig_NUVr_Xray}c).
Thereby, one can see how the threshold $HR=-0.2$, as used by \cite{Melnyk2013} to separate hard ($HR>-0.2$) and soft ($HR<-0.2$) X-ray dominated sources, may be relevant to identify the AGNs with heavy obscuration of their soft X-ray emission at $0.2 < z < 0.5$.  

In the following, the term "heavily-obscured X-ray AGN" always refers to an AGN with heavy obscuration of its soft X-ray emission and characterized by $HR>-0.2$.

\subsection{Type-2 AGN hosts along the $\Mstar$ quenching channel}
\label{sect_AGNs}

In Fig.~\ref{fig_AGN_NUVr}, we make use of such distinction between hard ($HR>-0.2$) and soft ($HR<-0.2$) X-ray dominated AGNs when showing the $\NUVr$ distribution of X-ray AGN host galaxies along the $\Mstar$ quenching channel together with radio-loud ones.
We show the AGN host galaxy number counts as a function of $\NUVr$ in the  top panel (Fig.~\ref{fig_AGN_NUVr}a), and the corresponding fraction of galaxies hosting an AGN in the  bottom panel (Fig.~\ref{fig_AGN_NUVr}b).

While Fig.~\ref{fig_AGN_NUVr} confirms radio-loud AGNs to be mostly concentrated in already quiescent galaxies (75 per cent of radio-loud AGNs are hosted by galaxies with sSFR $< 10^{-11} M_\odot \mathrm{yr}^{-1}/ M_\odot$), one can see how the fraction of galaxies hosting a radio-loud AGN increases with increasing $\NUVr$ colour (i.e., with decreasing sSFR) to finally become the most probable kind of AGN in the most quiescent galaxies with $\NUVr \sim 6$ (i.e., sSFR $< 10^{-14} M_\odot \mathrm{yr}^{-1}/ M_\odot$), $\sim$0.6 per cent of which host a radio-loud AGN.
To complete the picture, 75 per cent of radio-loud AGNs are hosted by galaxies with stellar mass $M_* >10^{11}M_\odot$, confirming radio-loud AGNs to be hosted by very massive quiescent galaxies, which we may expect to inhabit DM halos with masses of a few $10^{13}M_\odot$ \citep[assuming a stellar-to-halo-mass relation, e.g.,][]{Coupon2015, Legrand2018}. Our results are therefore in line with previous studies where radio-loud AGNs were also found to be highly clustered or to reside in rich environments \citep[e.g.,][]{Hickox2009, Malavasi2015}.

On the other hand, Fig.~\ref{fig_AGN_NUVr} confirms that X-ray AGN host galaxies globally follow the distribution of galaxies, with a peak of their incidence in the green valley, as observed previously for optically-obscured AGN host galaxies \citep[e.g.,][]{Hickox2009, Schawinski2014}.
More interestingly, our results do show that heavily-obscured X-ray AGNs are the very reason why the incidence of X-ray AGNs peaks in the green valley where their fraction outranks any other kind of optically-obscured AGN. 
The median fraction of ($HR>-0.2$) heavily-obscured X-ray AGNs is indeed observed to rise from less than 0.2 per cent in SF galaxies up to 0.8 per cent in the green valley, and then to decrease down to less than 0.2 per cent in Q galaxies.
Conversely, ($HR<-0.2$) unobscured X-ray AGNs prevail in SF and Q galaxies with a maximum of 0.4 per cent of host galaxies.
Consistently, the stellar mass of galaxies hosting heavily-obscured X-ray AGNs peaks around $\Mstar$, with a median mass of $M_* \simeq 10^{10.65}M_\odot$ and 60 per cent of these galaxies with $10^{10.4} \leq  M_*/M_\odot \leq 10^{10.9}$.

\section{Discussion}
\label{sect_discussion}

As described in Section~\ref{sect_results}, the NUVrK distributions of our samples of radio-loud and X-ray AGN host galaxies are quite different. In particular, the NUVrK colours of the galaxies hosting radio-loud and heavily-obscured X-ray AGNs suggest that these two kinds of AGNs could be associated with different stages of galaxy evolution.
In the following sections, we discuss the implications of these findings for the quenching scenario of $\Mstar$ galaxies.

\subsection{Radio-loud AGNs are mostly hosted by quiescent galaxies}

Figs~\ref{fig_NUVrK} and \ref{fig_AGN_NUVr} suggest that some galaxies develop a radio-loud AGN as they leave the green valley and move into the quiescent population. The presence of a radio-loud AGN, inefficiently accreting hot gas \citep[see, e.g.,][]{Hardcastle2007}, is often regarded as a viable mechanism to prevent further gas cooling, accretion and subsequent star-formation \citep[e.g., see][]{Croton2006}. Still, the probability of a galaxy developing a radio AGN at its core may depend on its mass \citep[with radio-loud AGNs preferentially found in more massive galaxies, see][and references therein]{Best2005, Bardelli2009} or on the environment in which it resides \citep[e.g.,][]{Malavasi2015, Bardelli2010}. This may explain why, while radio-loud AGNs are the most probable kind of AGN among quiescent galaxies, only a small fraction of these galaxies exhibit radio-loud AGNs (Fig.~\ref{fig_AGN_NUVr}b).

On the other hand, the fact that radio-loud AGNs are mostly concentrated in already quiescent galaxies tends to rule out radio-loud AGN feedback as a primary mechanism for the quenching of $\Mstar$ galaxies.

\subsection{Heavily-obscured X-ray AGNs as the final stages of major mergers?}

As detailed in Sect.~\ref{sect_AGNs}, our results argue for a mechanism that favours heavily-obscured X-ray AGNs (i.e., AGNs with heavy obscuration of their soft X-ray emission) in $\Mstar$ galaxies that are transitioning in the green valley (see Figs~\ref{fig_NUVr_Xray}a and b).
Heavy obscuration of soft X-ray emission is generally associated with obscuring material surrounding the supermassive black hole (SMBH) on fairly small scales -- e.g., material associated with the SMBH torus observed edge-on in the AGN \textit{unified model} \citep{Antonucci1993, UrryPadovani1995}. However, in such an orientation-based model, heavy obscuration should randomly affect the galaxy population and not a specific phase of the galaxy evolution or, in our case, a specific region of the NUVrK colour space as observed in Sect.~\ref{sect_AGNs}. 
Rather, our results argue for an evolutionary model where the heavy obscuration of soft X-rays is caused by a dramatic event like a merger \citep[for a review about obscured AGNs, see][]{HickoxAlexander2018}.

As a matter of fact, the soft X-ray obscuration of an AGN --as traced by its X-ray hardness ratio-- can be associated with a corresponding hydrogen column density, $N_H$, along the line of sight, and both models \citep[e.g.,][]{Hopkins2006, Blecha2018} and observations \citep[e.g.,][]{Satyapal2014, Kocevski2015, Ricci2017a} have highlighted the tight connection between galaxy mergers and very high column densities (i.e., typically $N_H \sim 10^{22-24} \mathrm{cm}^{-2}$). The maximum column density reached during the merger has been predicted to vary with the gas content of the galaxies and to decrease after the coalescence of the SMBHs \citep[][]{Blecha2018}. This is consistent with the decreasing obscuration (and conversely, increasing luminosity) of soft X-rays observed after the green valley (Figs~\ref{fig_NUVr_Xray}a and b). It may also be related to the expulsion of the circumnuclear material by the AGN's radiation pressure when reaching high accretion rates \citep[typically with Eddington ratios of $\lambda_\mathrm{Edd} \simeq 10^{-2}$;][]{Ricci2017b}.

In our case, the median hardness ratio $HR=0.26$ of green-valley $\Mstar$ galaxies is typical of column densities of $N_H \sim 3 \times 10^{22} \mathrm{cm}^{-2}$ at $0.2 < z < 0.5$ \citep[see, e.g.,][]{Hickox2007}. This value of column density is in very good agreement with what is predicted by high-resolution simulations in the final stage of major mergers of gas-poor galaxies with stellar masses of $M_* \sim 2.5-5 \times 10^{10}M_\odot$ \citep[][]{Blecha2018},
which is consistent with the decreasing gas content of quenching galaxies, in the starvation scenario put forth for $\Mstar$ galaxies \citep[][]{Schawinski2014, Peng2015, Moutard2016b}.
Moreover, the fraction of heavily-obscured ($HR>-0.2$) AGN host galaxies we observe in the green valley is consistent with the duration predicted for the late stage of $\Mstar$ galaxy mergers, namely, 10--150 Myr \citep[][]{Blecha2018}: Assuming $\Mstar$ galaxies spend 1--3.5 Gyr in the \textit{green valley} \citep[][]{Moutard2016b}, the fraction of heavily-obscured AGN host galaxies is expected to be between 0.3 and 15 per cent, which is consistent with the 0.4--0.8 per cent we found (see Fig.~\ref{fig_AGN_NUVr}b).

In summary, our key finding is that the incidence of heavily-obscured X-ray AGNs observed in $\Mstar$ galaxies \textit{in} the green valley suggests that mergers occur \textit{after} these quenching galaxies have left the SF main sequence.   
While mass-related mechanisms have been successfully put forth to explain the quenching of star formation in galaxies reaching the characteristic stellar mass $\Mstar$, such as virial shock heating within DM halos reaching a few $10^{12} M_\odot$ \citep[e.g.,][]{Keres2005, DekelBirnboim2006} of star formation feedback \citep[e.g.,][]{Peng2010}, these models alone would fail to explain an increasing incidence of mergers \textit{in} the green valley, i.e., \textit{after} the onset of the quenching process. Instead, our observations support models in which merging occurs after the quenching galaxy has left the main sequence, and in the next section we discuss how these observations seem to fit quite naturally within the theoretical framework developed by \citet{Pichon2011} and \citet{Codis2015}.

\subsection{Proposed framework for $\Mstar$ galaxy quenching}
\label{sect_scenar}

In the  \citet{Pichon2011} and \citet{Codis2015} framework, galaxies tend to be formed within DM halos whose spins are aligned with their closest cosmic filaments \citep{Pichon2011}, due to cold gas streams flowing toward those filaments. Eventually, DM halos experience a flip of their angular momentum due to (major) mergers with other DM halos \citep{Welker2014} as they fall along cosmic filaments in the direction of clusters. This flip of the angular momentum is precisely predicted to occur for DM halos reaching masses of a few $10^{12} M_\odot$ \citep[][]{Codis2012, Codis2015} which is, as derived from the stellar-to-halo mass relation \citep[e.g.][]{Coupon2015, Legrand2018}, the mass of the halos that host the $\Mstar (\simeq 10^{10.6}M_\odot)$ galaxies we are interested in, on average. 
This picture is consistent with the outputs of hydrodynamical simulations in which galaxies are also predicted to eventually experience a spin flip when reaching $M_* \sim 10^{10.1-10.5}M_\odot$ \citep[][]{Dubois2014, Codis2018}.

In other words, in this theoretical framework, mergers of $\sim 10^{12} M_\odot$ DM halos --and mergers of the $\sim\Mstar$ galaxies that are hosted by these halos-- are favoured along cosmic filaments.  The coalescence of $\sim\Mstar$ galaxy SMBHs, which are expected to occur in the final stages of the galaxy-galaxy mergers,  are in line with this framework. 
Our results, which support a picture where the coalescence of $\sim\Mstar$ galaxy SMBHs happens in the \textit{green valley} (i.e., after the quenching has started) are therefore in line with a scenario where $\Mstar$ galaxy quenching happens along cosmic filaments.  
We note that this is consistent with observations that show that, on average, $\sim\Mstar$ quiescent galaxies are found to be closer to cosmic filaments than their SF counterparts \citep[e.g.,][]{Malavasi2017, Laigle2018, Kraljic2018}, i.e., that the quenching probability of $\sim\Mstar$ galaxies increases with decreasing distance to filaments.

The precise mechanisms that disconnect  $\sim 10^{12} M_\odot$ DM halos from cold-gas streams around filaments are not yet fully understood, but may involve simultaneously mass (virial shock heating) and/or position and size of the halos within filaments \citep[][]{Laigle2015}. At present we cannot rule out either a significant effect of the interactions between galaxy/DM halo angular momentum and cold-gas streams \citep{Welker2014, AragonCalvo2016}, star formation having been found to be lower in $\gtrsim\Mstar$ galaxies when they are in pairs \citep[e.g.,][]{Coenda2019}, which may suggest the quenching of $\sim\Mstar$ galaxies to coincide more or less with the merger of their DM host halos.

In any case, beyond being compatible with our observations, the interest of the picture we just drawn for $\Mstar$ galaxy quenching is to be easily falsifiable, through further observations and analyses. 
In particular, the relevance of the scenario we discussed will be tested by answering the following questions:
\begin{itemize}
\item[1)]  Is the morphology of heavily-obscured, $HR>-0.2$ X-ray AGN-host galaxies compatible with final stages of galaxy mergers?
\item[2)]  What is the typical distance of those AGN-host galaxies to the closest cosmic filament, with respect to other galaxies?
\item[3)]  What is the fraction of pairs of $\Mstar$ galaxies in the green valley, compared to the fraction of pairs of $\Mstar$ galaxies outside the green valley (within the quiescent and star formation main sequences)?
\item[4)]  What is the (typical) dark-matter halo mass of $\Mstar$ galaxies in the green valley, compared to $\Mstar$ galaxies outside the green valley?
\end{itemize}
Our intent is to tackle those questions in future analyses.

\section{Summary}

In the present paper, we analysed the connection between X-ray and radio-loud (optically-obscured) AGNs and the physical properties of their evolved and massive host galaxies at $0.2 \leq z \leq 0.5$. More specifically, we examined the interplay between the mass-related quenching channel followed by $\Mstar (\simeq 10^{10.6}M_\odot)$ galaxies in the rest-frame NUVrK colour diagram, and hosted AGN properties.

Our results confirm that (1) radio-loud AGNs are mostly hosted by already-quenched galaxies \citep[Figs~\ref{fig_NUVrK} and  \ref{fig_AGN_NUVr}; see also][]{Hickox2009}, tending to rule out radio-loud AGN feedback as the primary reason for $\Mstar$ galaxy quenching.
Furthermore, our analysis shows that (2) the AGNs suffering from heavy obscuration of their soft X-ray emission are preferentially hosted by $\Mstar$ galaxies that are in the NUVrK green valley (i.e., $\Mstar$ galaxies with intermediate sSFR; Figs~\ref{fig_NUVrK} and  \ref{fig_AGN_NUVr}), with a median hardness ratio of $HR = 0.26$ corresponding to column densities of $N_H \sim 3 \times 10^{22} \mathrm{cm}^{-2}$, which is typically what is predicted for gas-poor $\Mstar$-galaxy major mergers \citep{Blecha2018}.
Thereby, our results may support a quenching scenario where (gas-poor) $\Mstar$ galaxies experience major mergers \textit{after} they have left the SF main sequence, i.e., a scenario where galaxy mergers are not the very cause but rather an aftermath of the quenching mechanism(s).

Whilst purely mass-related mechanisms that have been put forth to explain the quenching of $\Mstar$ galaxies \citep[e.g.][]{Keres2005, Peng2010} would fail to account for a increasing incidence of major mergers for $\Mstar$ galaxies that are in the process of quenching, 
it is interesting to note that this falls quite naturally within the theoretical framework developed by \citet{Pichon2011} and \citet{Codis2015}. 
In this framework, dark-matter halos reaching a mass of $\sim 10^{12}M_\odot$ \citep[i.e., the typical hosts of $\Mstar$ galaxies;][]{Coupon2015, Legrand2018} are predicted to experience a merger-induced flip of their spin \citep{Welker2014}, as they fall along cosmic filaments.
In other words, our results invite us to consider a picture where $\Mstar$ galaxy quenching may preferentially happen along cosmic filaments.

In summary, \textit{mass quenching} ---as embodied, by definition, by the quenching of $\Mstar$ galaxies at $z < 1.5$--- may be more complicated than a purely mass-related process.
While the quenching of $\Mstar$ galaxies is expected to be slow and compatible with \textit{starvation} processes \citep[][]{Schawinski2014, Peng2015, Moutard2016b}, the precise mechanism(s) disconnecting their $\sim 10^{12}M_\odot$ host DM halos from cold-gas streams remain(s) unclear and may have multiple causes, from virial shock heating within DM halos \citep[e.g.,][]{DekelBirnboim2006}, to vorticity within filaments \citep{Laigle2015}, which may potentially be affected by major mergers between halos along filaments \citep{Welker2014, AragonCalvo2016}.

The pilot study we present here is an invitation to confirm and refine the scenario of the quenching of $\Mstar$ galaxies.
In particular, morphological analysis from much deeper imaging will allow us to increase drastically the number of $\Mstar$ galaxy mergers and pairs observed in the green valley, aiming to investigate the role of DM halo mergers in a future paper, while the connection between $\Mstar$ green-valley galaxies and cosmic filaments will be specifically investigated in a companion study.

\section*{Acknowledgements}

We gratefully acknowledge the anonymous reviewer, whose insightful comments helped in improving the clarity of the paper.
T.~M. wishes to thank Luigi C. Gallo and Ivana Damjanov for useful discussions.
This research was supported by the ANR Spin(e) project (ANR-13-BS05-0005, \texttt{http://cosmicorigin.org}) and by a Discovery Grant from the Natural Sciences and Engineering Research Council (NSERC) of Canada. 
This research has been supported by the funding for the ByoPiC project from the European Research Council (ERC) under the European Union's Horizon 2020 research and innovation programme grant agreement ERC-2015-AdG 695561.
This research makes use of the VIPERS-MLS database, operated at CeSAM/LAM, Marseille, France. This work is based in part on observations obtained with WIRCam, a joint project of CFHT, Taiwan, Korea, Canada and France. The CFHT is operated by the National Research Council (NRC) of Canada, the Institut National des Science de l'Univers of the Centre National de la Recherche Scientifique (CNRS) of France, and the University of Hawaii. This work is based in part on observations made with the Galaxy Evolution Explorer (GALEX). GALEX is a NASA Small Explorer, whose mission was developed in cooperation with the Centre National d'Etudes Spatiales (CNES) of France and the Korean Ministry of Science and Technology. GALEX is operated for NASA by the California Institute of Technology under NASA contract NAS5-98034. This work is based in part on data products produced at TERAPIX available at the Canadian Astronomy Data Centre as part of the Canada-France-Hawaii Telescope Legacy Survey, a collaborative project of NRC and CNRS. The TERAPIX team has performed the reduction of all the WIRCAM images and the preparation of the catalogues matched with the T0007 CFHTLS data release.
Based on data products from observations made with ESO Telescopes at the La Silla Paranal Observatory as part of the VISTA Deep Extragalactic Observations (VIDEO) survey, under programme ID 179.A-2006 (PI: Jarvis).




\bibliographystyle{mnras}
\bibliography{Moutard2019_revised.bib} 






%


\bsp	
\label{lastpage}

\end{document}